# A BCS-GDE Multi-objective Optimization Algorithm for Combined Cooling, Heating and Power Model with Decision Strategies

Jiaze Sun [a,b,c], Jiahui Deng [d *], Yang Li [e], Nan Han [d]

Corresponding author email: dengjiahui@stumail.nwu.edu.cn (Jiahui Deng)

[a] *School of Computer Science and Technology, Xi'an University of Posts and Telecommunications, Xi'an 710121, China*

[b] *Shaanxi Key Laboratory of Network Data Analysis and Intelligent Processing, Xi'an 710121, China*

[c] *Xi 'an Key Laboratory of Big Data and Intelligent Computing, Xi'an 710127, China*

[d] *School of Information Science and Technology, Northwest University, Xi'an 710127, China*

[e] *School of Electrical Engineering*, *Northeast Electric Power University*, *Jilin 132012, China*

## ABSTRACT

District energy systems can not only reduce energy consumption but also set energy supply dispatching schemes according to demand. In addition to economic cost, energy consumption and pollutant are more worthy of attention when evaluating combined cooling, heating and power (CCHP) models. In this paper, the CCHP model is established with the objective of economic cost, primary energy consumption, and pollutant emissions. The mathematical expression of the CCHP system is proposed, and a multi-objective optimization model with constraints is established. According to different usage requirements, two decision-making strategies are designed, which can adapt to different scenarios. Besides, a generalized differential evolution with the best compromise solution processing mechanism (BCS-GDE) algorithm is proposed to optimize the CCHP model for the first time. The algorithm provides the optimal energy scheduling scheme by optimizing the production capacity of different capacity equipment. The simulation is conducted in three application scenarios: hotels, offices, and residential buildings. The simulation results show that the model established in this paper can reduce economic cost by 72%, primary energy consumption by 73%, and pollutant emission by 88%. Concurrently, the Wilcoxon signed-rank test shows that BCS-GDE is significantly better than OMOPSO, NSGA-II, and SPEA2 with greater than 95% confidence.

# 1   Introduction

## 1.1   Background information

Combined cooling heating and power (CCHP) system is used to provide distributed energy usually, it provides the energy required by the buildings. The CCHP system can recover the waste heat of power generation effectively, also can reduce the energy loss in transmission, and improve the energy supply efficiency [1]. The conversion rate of fuel to available energy is more than 80%. In the past related work, most of the CCHP problems were modeled for the single-objective of economic [2], but since there are many influencing factors and issues that need to be considered, combined cooling heating and power economic emission dispatch (CCHPEED) has been paid more attention. In the current study, a series of multi-objective optimization models have been proposed according to different regions and energy demands. In view of this problem, how to establish a reasonable decision-making strategy model according to the actual situation, and how to find the optimal resource dispatching scheme has become the current research focus.

In the face of energy scheduling, most of the existing studies are based on fixed energy demand and parameter influence models, but in practice, the energy demands of different seasons or buildings are different. The parameters, for example, electricity price, in different periods of a whole day are also different, which requires dynamic adjustment of the structure and parameters of the model for energy scheduling optimization. Besides, in the current research, although evolutionary algorithms are used to solve the CCHPEED, most of the multi-objective problems are transformed into the single-objective problem in the process of problem processing in the references, and the non-dominated solution is not considered, the solution may not be the optimal solution of the model. While how to evaluate the optimization performance of the multi-objective evolutionary algorithm, there is no clear method and simulation experiment in the research papers.

## 1.2   Literature survey

In related work, most of the research on CCHP and CHP systems are based on single-objective, namely economic cost, and the goal is to find the best dispatching scheme under the optimal economic situation [3] [4] [5] [6] [7] [8] [9]. In recent years, the multi-objective CCHP system research has also been widely concerned by scholars. Tezer et al. [10] solved the problems of minimizing energy costs and carbon dioxide emissions in hybrid systems. Aiming at investment in distributed heating and power supply systems, a multi-objective optimization model was proposed [11], which uses to minimize the total economic cost and carbon dioxide emissions, the model was applied to the city to analyze the multi-objective problems and solutions. Chen et al. [12] designed a CCHP system with efficiency and cost as objective functions. Validated the effectiveness of the system under different conditions. Dorotić et al. [13] proposed an hour-based multi-objective optimization district heating and cooling model, which can define the supply capacity including heat storage capacity and its annual operation. An optimization model with three objectives including economic cost, primary energy consumption, and pollutant emissions was established and applied in five cities, respectively [14], and the optimal operation scheme algorithm was used to determine the optimal operation mode of the model. Hu et al. [15] proposed a stochastic multi-objective optimization model, the probability constraints were added to the stochastic model to optimize CCHP operation strategies under different climatic conditions. Dorotić et al. [16] carried out a multi-objective optimization of the district heating system. The model can optimize the hourly running time of the annual time range and optimize the scale of supply capacity including storage. Cao et al. [17] developed a mathematical model to analyze a novel ejector enhanced heat pump system from the thermodynamics viewpoint, the NSGA-II was applied to solve the multi-objective optimization model. However, experiments do not consider verification in real scenarios.

Although numerous models have been proposed to optimize resource scheduling, few works analyzed the demands and price of primary energy in different periods of a whole day, this is not very practical. The existing literature rarely provides different system models for different conditions, the suitability is not verified.

At present, there are many methods used to solve multi-objective problems, mainly divided into mathematical methods and heuristic algorithms. Mathematical methods are commonly used in linear programming (LP), mixed integer linear programming (MILP) [18], and nonlinear mixed integer linear programming (MINLP) [19] [20]. For some complex multi-objective optimization problems, the heuristic algorithm can solve them more effectively. For example, the impacts of the different energy price policies on the configuration of CHP and CCHP systems were studied by Tichi et al. [21] using the particle swarm optimization (PSO) algorithm. Meanwhile, a great quantity of multi-objective evolutionary algorithms was used to solve CHP models [22] [23] [24] [25] [26]. For CCHP problems, PSO and genetic algorithm (GA) are applied to optimize the models by literature [27] [28] [29]. A multi-objective optimization model was established by Wei et al. [30] to maximize the energy saving rate and minimize the energy expenditure, and the NSGA-II algorithm was used to determine a series of optimal resource scheduling strategies. Several excellent multi-objective optimization algorithms have also been proposed in recent years to address the schedule problems, for example, Cao et al. [31] focused on a 4E analyses of a biomass gasifier integrated energy system, using multi-objective bat optimization. There are also studies on load prediction and parameter optimization using the evolutionary algorithm and neural network [32] [33].

In the above literature, most of them used the weight function to convert multi-objective into single-objective models when performing multi-objective calculations. The results of the simulation do not reflect the unique nature of multi-objective problems [13] [16] [34]. While the existence of a non-dominated solution can provide different resource scheduling options. Most of the algorithms have no specific indicator to measure the quality of the optimization and the performance of the algorithm.

### 1.3 The work of this paper

In this paper, a CCHPEED model is established, which has three objectives, namely, economic cost, primary energy consumption (PEC), and carbon dioxide emissions (CDE). Three objectives need to be optimized concurrently, and a series of constraints are established. For the CCHPEED, the authors put forward a generalized difference algorithm with the best compromise solution processing mechanism (BCS-GDE) to optimize the dispatching scheme. The paper also employs the evaluation method of the multi-objective algorithm to evaluate BCS-GDE. To verify the effectiveness of the algorithm and the model, the other two decision-making models are used in the energy allocation of three different types of buildings respectively in the simulation part, and the model is applied to three different scenarios to verify the applicability.

The main contributions of this paper:

1. A multi-objective CCHPEED model with constraints is established, and two decision-making strategies are proposed, which can optimize the resource scheduling scheme dynamically;

2. For the CCHPEED problem, a BCS-GDE algorithm is proposed for the first time, which can select the approximate optimal solution in the optimal solution set and provide the optimal dispatching scheme as needed;

3. Simulation is conducted in three different types of buildings and three different scenarios. The simulation results are evaluated by a multi-objective performance evaluation indicator, and the three-dimensional images of Pareto solution set are drawn. The simulation results are compared with the classical evolutionary algorithms.

### 1.4 Organization of this paper

The arrangement of the remaining sections in this article is as follows: Section 2 establishes three models of the CCHP system under different conditions. Section 3 describes the multi-objective evolutionary algorithm proposed in this article and details the steps and processes. Section 4 provides simulation data and scene introduction. Section 5 discusses the simulation results, analyzes the system usage and optimization results in different situations. Section 6 is a summary of the article.

## 2 Formulation of the CCHP model

In this section, a CCHP model is established. The combined cooling, heating and power demand of the system are provided by the Power Grid (PG), power generation units (PGU), and boilers. The fuel is supplied to the generator, which produces power and waste heat. The power can be used by the building directly, also, the power demand can be met by purchasing from the grid. The recovered waste heat is used for refrigeration or heating to meet the cooling and heating demands of the buildings, and the boilers can be used to support the demand too. The schematic of a CCHP system is shown in Fig.1.

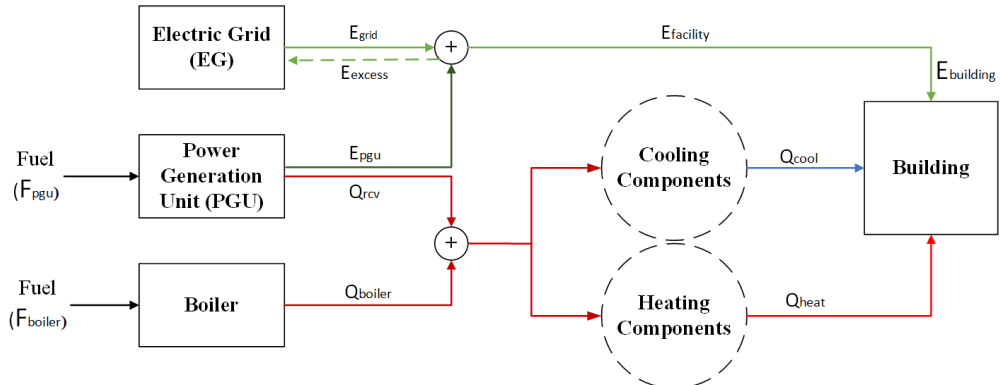

Fig 1 The schematic of a CCHP system

Fig.2 is a network model based on the energy flow of the CCHP system. Nodes 9, 10, and 11($E_d$, $Q_{c\_d}$, $Q_{h\_d}$) in the figure are the power, cooling, and heat demands of the system. Node 1 is a concept node, which represents the total energy required by the system.

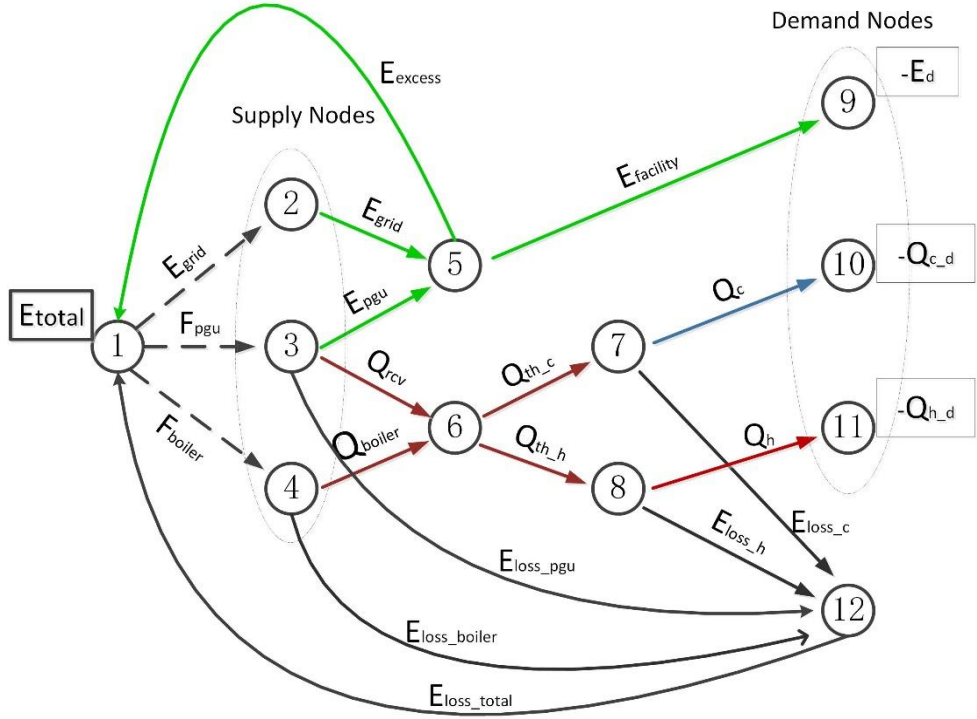

Node 1: Total energy required

Node 2: Power Grid

Node 3: PGU

Node 4: Boiler

Node 5: Electric energy provided by PGU and power grid

Node 6: Thermal energy provided by PGU and boiler

Node 7: CHP cooling components

Node 8: CHP heating components

Node 9: Electric energy demand

Node 10: Cooling energy demand

Node 11: Heating energy demand

Node 12: Total energy loss

Fig 2 Network flow model of a typical CCHP model

### 2.1 Objective functions

In the CCHP model built in this section, we consider three objectives. The first objective ($f_{cost}$) is to minimize the total cost of the system in T periods; the second objective ($f_{PEC}$) is to minimize the primary energy consumption in T periods; the third objective ($f_{CDE}$) is to minimize the carbon dioxide emissions of the system in T periods. The objective functions are as follow [14]:

$$Min \ f_{cost} = \sum_{t=1}^{T}\sum_{i}(C_{fuel,i,t})$$

$$= \sum_{t=1}^{T}\{C_{el}E_{grid}(t) + C_{f\_pgu}E_{pgu}(t) + C_{f\_boiler}Q_{boiler}(t)/\eta_{boiler}\} \quad (1)$$

$$Min\ f_{PEC} = \sum_{t=1}^{T}\sum_{i}(F_{fuel,i,t})$$

$$= \sum_{t=1}^{T}\left\{ECF_{PEC}E_{grid}(t) + FCF_{PEC_{pgu}}(a*E_{pgu}(t)+b) + FCF_{PEC_{boiler}}Q_{boiler}(t)/\eta_{boiler}\right\} \quad (2)$$

$$Min\ f_{CDE} = \sum_{t=1}^{T}\sum_{i}(E_{co_2,i,t})$$

$$= \sum_{t=1}^{T}\left\{ECF_{CDE}E_{grid}(t) + FCF_{CDE\_pgu}E_{pgu}(t) + FCF_{CDE\_boiler}Q_{boiler}(t)/\eta_{boiler}\right\} \quad (3)$$

where $t \in \{1,2,\dots,T\}$; in the objective of minimizing costs, the cost mainly consists of purchasing power from the power grid, PGU energy consumption cost, and boiler energy consumption cost. $C_{el}$, $C_{f_{pgu}}$ and $C_{f_{boiler}}$ respectively represent the cost of purchasing 1kwh electricity, the fuel cost of generating 1kwh energy in PGU, and the fuel cost of generating 1kwh energy in the boiler; $\eta_{boiler}$ represents the energy conversion efficiency of the boiler. In the objective of minimizing primary energy consumption, the energy consumption mainly consists of the power purchased by the grid, the energy consumed by PGU, and the energy consumed by the boiler. $ECF_{PEC}$, $FCF_{PEC\_pgu}$ and $FCF_{PEC\_boiler}$ refer to the conversion factor of primary energy for purchasing power, the primary energy conversion factor of fuel used in PGU, and the primary energy conversion factor of fuel used in the boiler; $a$ and $b$ are fuel electric energy conversion parameters. In the objective of minimizing carbon dioxide emissions, the emission mainly consists of the emission of pollutants from the power grid, the emission of pollutants from PGU conversion fuel, and the emission of pollutants from boiler combustion. $ECF_{CDE}$, $FCF_{CDE\_pgu}$ and $FCF_{CDE\_boiler}$ represent the carbon dioxide emission conversion factor of power, the emission conversion factor of PGU fuel, and the emission conversion factor of boiler fuel, respectively.

### 2.2 Constraints

In this model, a series of constraints need to be met to optimize the objective functions, including energy conservation constraints, fuel conversion constraints, and energy efficiency constraints. The constraint formula is as follows.

The energy conservation constraints of nodes 3, 4, 5, 6, 7, 8, and 12 in Fig.2 are as follows [14]. Equations (4) and (5) are the conservation conditions of converting the energy generated by PGU and boiler to available energy, respectively. The energy flow diagram of PGU and boiler is shown in Fig.3.

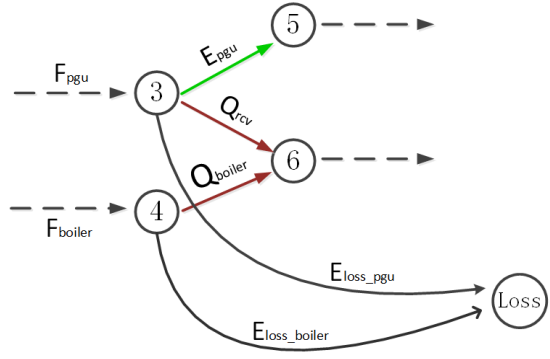

Fig 3 The energy flow diagram of nodes 3 and 4

$$E_{pgu}(t) + Q_{rcv}(t) + Energy_{loss\_pgu}(t) - F_{pgu}(t) = 0 \qquad (4)$$

$$Q_{boiler}(t) + Energy_{loss\_boiler}(t) - F_{boiler}(t) = 0 \qquad (5)$$

Equations (6) and (7) respectively represent the conservation relationship between the total power generated, the total heat generated, and the energy output. The energy conservation relationship of nodes 5 and 6 are shown in Fig.4.

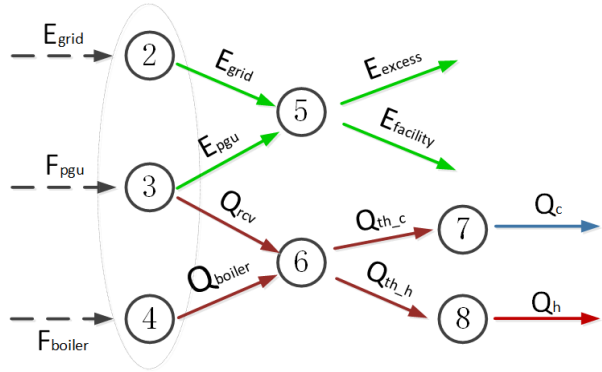

Fig 4 The energy conservation relationship of nodes 5 and 6

$$E_{excess}(t) + E_{facility}(t) - E_{grid}(t) - E_{pgu}(t) = 0 \qquad (6)$$

$$Q_{th\_cool}(t) + Q_{th\_heat}(t) - Q_{rcv}(t) - Q_{boiler}(t) = 0 \qquad (7)$$

Formulas (8) and (9) respectively represent the conservation relationship between the energy generated by the cooling component, the heating component, and the output energy. The schematic diagram is shown in Fig.5.

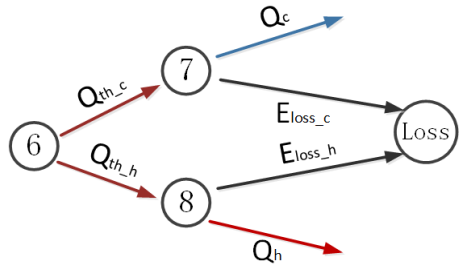

Fig 5 The energy conservation relationship of nodes 7 and 8

$$Q_{cool}(t) + Energy_{loss\_c}(t) - Q_{th\_cool}(t) = 0 \tag{8}$$

$$Q_{heat}(t) + Energy_{loss\_h}(t) - Q_{th\_heat}(t) = 0 \tag{9}$$

Equation (10) shows the conservation relation of the total energy loss of the system. Each component generates energy losses during operation, and the balance of losses is also one of the constraints that need to be observed. The relationship diagram of energy loss is shown in Fig.6.

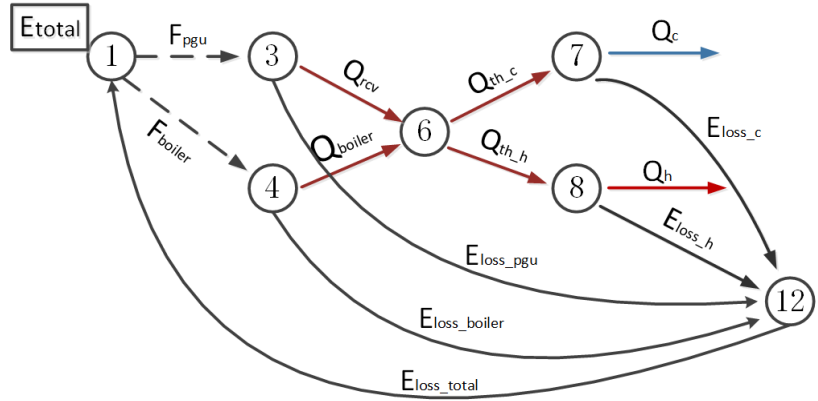

Fig 6 Energy loss flow diagram

$$Energy_{loss\_total}(t) - Energy_{loss\_pgu}(t) - Energy_{loss\_boiler}(t) - Energy_{loss\_c}(t) - Energy_{loss\_h}(t) = 0 \tag{10}$$

The linear relationship between PGU fuel and electric energy is shown in Fig.7 [14], and the relationship expression is shown in equation (11), which is a typical illustration:

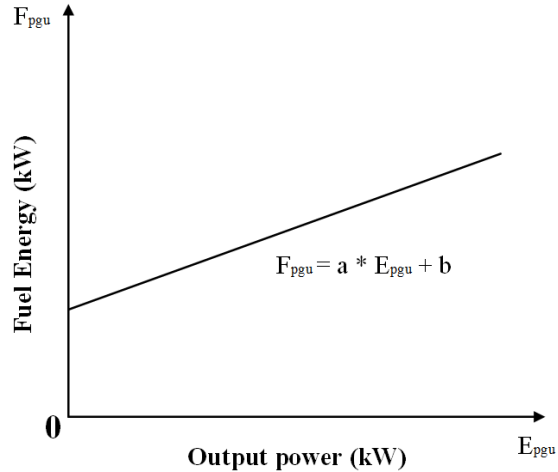

Fig 7 Linear relation between fuel energy and power output

$$F_{pgu}(t) = \begin{cases} a * E_{pgu}(t) + b & E_{pgu}(t) > 0 \\ 0 & E_{pgu}(t) = 0 \end{cases} \tag{11}$$

where $a$ and $b$ are the conversion parameters of PGU fuel to power. PGU, boiler, cooling component, and heating component have energy conversion efficiency in the energy supply system, and their conversion efficiency constraints are shown in formula (12-15) [14], respectively:

$$Q_{rcv}(t) - \eta_{pgu\_th}F_{pgu}(t) = 0 \tag{12}$$

$$Q_{boiler}(t) - \eta_{boiler}F_{boiler}(t) = 0 \tag{13}$$

$$Q_{cool}(t) - \eta_{cool\_comp}Q_{th\_cool}(t) = 0 \tag{14}$$

$$Q_{heat}(t) - \eta_{heat\_comp}Q_{th\_heat}(t) = 0 \tag{15}$$

where, $\eta_{pgu\_th}$, $\eta_{boiler}$, $\eta_{cool\_comp}$ and $\eta_{heat\_comp}$ is the conversion efficiency of PGU, boiler, cooling component, and heating component, respectively.

## 2.3 Decision strategies for the CCHP model

In different seasons, there have different energy demands. For example, in summer, it needs refrigeration instead of heating; in winter, there is no need for refrigeration, but adequate heating is needed. The CCHP system is modified to meet different types of energy demand in different seasons. In sections 2.1 and 2.2, the model with both PGU and boiler is established, in this section, the other two different CCHP models are proposed.

### 2.3.1 Case1: PGU is off

In this model, PGU will be shut down, all the power demand by the system will come from the power grid, and all the cold and heat demand of the system will be supplied by the boiler. The schematic diagram is shown in Fig.8.

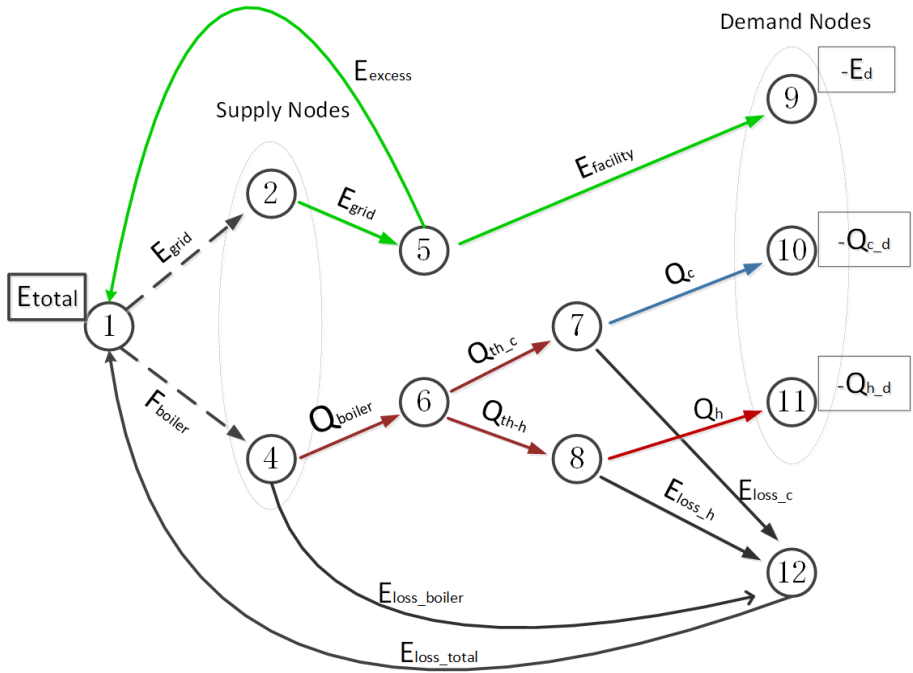

Fig 8 Network flow model of a CCHP model without PGU

When PGU is closed, the objectives established according to the model are changed. The impact of PGU is removed based on the original objectives. The objective functions of the model are established as follows [14]:

$$Min \ f_{cost} = \sum_{t=1}^{T}\sum_{i}(C_{fuel,i,t}) = \sum_{t=1}^{T}\{C_{el}E_{grid}(t) + C_{f\_boiler}Q_{boiler}(t)/\eta_{boiler}\} \quad (16)$$

$$Min \ f_{PEC} = \sum_{t=1}^{T}\sum_{i}(F_{fuel,i,t}) = \sum_{t=1}^{T}\{ECF_{PEC}E_{grid}(t) + Q_{boiler}(t)/\eta_{boiler}\} \quad (17)$$

$$Min \ f_{CDE} = \sum_{t=1}^{T}\sum_{i}(E_{co_2,i,t}) = \sum_{t=1}^{T}\{ECF_{CDE}E_{grid}(t)+FCF_{CDE\_boiler}Q_{boiler}(t)/\eta_{boiler}\} \quad (18)$$

In this case, the PGU is turned off. In constraints, the energy flow generated by the PGU is canceled. The constraints of energy conservation, fuel conversion, and energy efficiency are as follows [14]:

$$Q_{boiler}(t) + Energy_{loss\_boiler}(t) - F_{boiler}(t) = 0 \quad (19)$$

$$E_{excess}(t) + E_{facility}(t) - E_{grid}(t) = 0 \quad (20)$$

$$Q_{th\_cool}(t) + Q_{th\_heat}(t) - Q_{boiler}(t) = 0 \quad (21)$$

$$Q_{cool}(t) + Energy_{loss\_c}(t) - Q_{th\_cool}(t) = 0 \quad (22)$$

$$Q_{heat}(t) + Energy_{loss\_h}(t) - Q_{th\_heat}(t) = 0 \quad (23)$$

$$Energy_{loss\_total}(t) - Energy_{loss_{boiler}}(t) - Energy_{loss_c}(t) - Energy_{loss_h}(t) = 0 \quad (24)$$

$$Q_{boiler}(t) - \eta_{boiler}F_{boiler}(t) = 0 \quad (25)$$

$$Q_{cool}(t) - \eta_{cool\_comp}Q_{th\_cool}(t) = 0 \quad (26)$$

$$Q_{heat}(t) - \eta_{heat\_comp}Q_{th\_heat}(t) = 0 \quad (27)$$

### 2.3.2   Case 2: Boiler is off

When the boiler is shut down, the power demand of the system is provided by the grid and PGU jointly, while the heating and cooling demand of the system are all provided by PGU. The schematic diagram is shown in Fig.9.

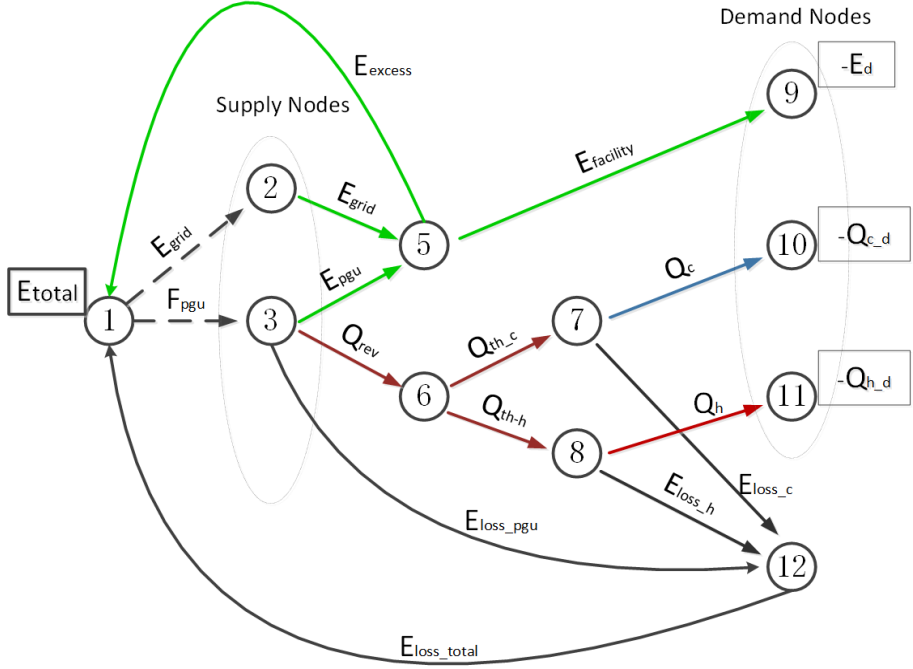

Fig 9 Network flow model of a CCHP model without boiler

When the boiler equipment is shut down, the objective functions of the model have also modified accordingly, and the influence of the boiler is removed based on the original model. The objective functions of the CCHP system are as follows [14]:

$$Min \ f_{cost} = \sum_{t=1}^{T}\sum_{i}(C_{fuel,i,t}) = \sum_{t=1}^{T}\{C_{el}E_{grid}(t) + C_{f\_pgu}E_{pgu}(t)\} \tag{28}$$

$$Min \ f_{PEC} = \sum_{t=1}^{T}\sum_{i}(F_{fuel,i,t}) = \sum_{t=1}^{T}\{ECF_{PEC}E_{grid}(t) + FCF_{PEC_{pgu}}(a*E_{pgu}(t) + b)\} \tag{29}$$

$$Min \ f_{CDE} = \sum_{t=1}^{T}\sum_{i}(E_{co_2,i,t}) = \sum_{t=1}^{T}\{ECF_{CDE}E_{grid}(t) + FCF_{CDE\_pgu}E_{pgu}(t)\} \tag{30}$$

The energy generation of the boiler is turned off in this case, and the constraint formula is also deleted. The constraints of the model are as follows [14]:

$$E_{pgu}(t) + Q_{rcv}(t) + Energy_{loss\_pgu}(t) - F_{pgu}(t) = 0 \tag{31}$$

$$E_{excess}(t) + E_{facility}(t) - E_{grid}(t) - E_{pgu}(t) = 0 \tag{32}$$

$$Q_{th\_cool}(t) + Q_{th\_heat}(t) - Q_{rcv}(t) = 0 \tag{33}$$

$$Q_{cool}(t) + Energy_{loss\_c}(t) - Q_{th\_cool}(t) = 0 \tag{34}$$

$$Q_{heat}(t) + Energy_{loss\_h}(t) - Q_{th\_heat}(t) = 0 \tag{35}$$

$$Energy_{loss\_total}(t) - Energy_{loss\_pgu}(t) - Energy_{loss\_c}(t) - Energy_{loss\_h}(t) = 0 \tag{36}$$

$$F_{pgu}(t) = \begin{cases} a * E_{pgu}(t) + b & E_{pgu}(t) > 0 \\ 0 & E_{pgu}(t) = 0 \end{cases} \tag{37}$$

$$Q_{rcv}(t) - \eta_{pgu\_th}F_{pgu}(t) = 0 \tag{38}$$

$$Q_{cool}(t) - \eta_{cool\_comp}Q_{th\_cool}(t) = 0 \tag{39}$$

$$Q_{heat}(t) - \eta_{heat\_comp}Q_{th\_heat}(t) = 0 \tag{40}$$

## 3 Optimization method

In Section 2, the optimization model of CCHP is established, three objectives are considered: economic cost, energy consumption, and pollutant emission. There are constraints such as conservation laws in energy transmission that need to be obeyed. To sum up, this can be summarized as a constrained multi-objective optimization problem in mathematics. In this section, the authors conduct energy scheduling by optimizing the values of three decision variables to minimize the objective function while satisfying the constraints. The decision variables are listed in Table 1.

Table 1 Decision variables of the models.

| Decision variables | Description |
| --- | --- |
| **X1** | Electricity purchased from the grid |
| **X2** | Natural gas consumption by PGU |
| **X3** | Natural gas consumption by the boiler |

Due to the mutual constraints between multiple objective functions, the decrease of one objective value will often lead to the increase of another objective value, The result of optimization cannot be determined. The algorithm proposed in this paper uses the idea of Euclidean distance to select the best compromise solution from the approximate optimal solution set, which solves this problem well.

Fig.10 is the overall flow chart of using the generalized differential evolution with the best compromise solution processing mechanism (BCS-GDE) algorithm to optimize resource scheduling. In this section, a multi-objective evolutionary algorithm is used to optimize the decision variables of the model. After obtaining a set of non-dominated solutions, the optimal compromise solution selection strategy is used to select the most ideal variable value, that is, the resource scheduling scheme of the CCHP model. The following subsections are the detailed algorithm process.

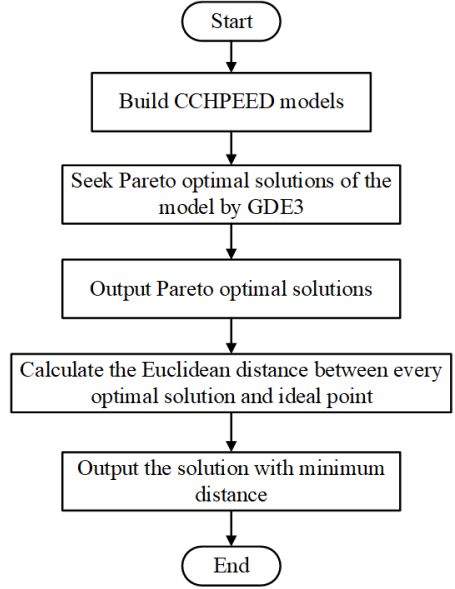

Fig 10 Overall flow chart of the optimization process

### 3.1 BCS-GDE

Since the model proposed in this paper is a multi-objective optimization problem with constraints, the generalized differential evolution with the best compromise solution processing mechanism (BCS-GDE) algorithm will be proposed to solve it.

BCS-GDE is an extension of differential evolution [35] (DE) and the third evolution step of generalized differential evolution [36] (GDE3). DE can only be used to solve single-objective optimization problems, while BCS-GDE can be used to solve multi-objective optimization problems with constraints. When there is only one objective function, the BCS-GDE algorithm falls back to DE algorithm for problem solving. BCS-GDE is different from GDE3 in that BCS-GDE can find the best compromise in the optimal solution set according to the rules in section 3.2. The following will introduce the detailed calculation process of BCS-GDE:

Step (1). Initialization: Create a parent population P, and set the value of population size to N. The initial values of individuals can be defined by (rand(0,1) $* (x_j^U - x_j^L) + x_j^L$), where $x_j^U$ and $x_j^L$ represent the upper and lower bounds of the individual x in the jth objective, respectively. The number of maximum iterations is set to $G_{max}$, and the generation counter $G = 0$.

Step (2). Selection: The selection operator is used to select three individuals in the population P randomly as parents, and they are different from each other and all are different from $x_i$.

Step (3). Mutation: Create an offspring population S with population size being 2*N and with no individual in it; perform differential mutation operation on individuals in population P. For the three selected parent individuals in Step (2), the vector difference of them is scaled first, and then vector synthesis is performed with another individual, that is, $U_{j,i,G} = x_{j,r_3,G} + F * (x_{j,r_1,G} - x_{j,r_2,G})$, where $i \neq r_1 \neq r_2 \neq r_3$. The parameter $F$ is a mutation operator, the size of $F$ will affect the diversity and convergence rate of the population. In this paper, $F = 0.5$ is set.

Step (4). Crossover: Differential crossover has been used to do crossover between $U_{i,G}$ and $x_{i,G}$. $U_{i,G}$ has been obtained from Step (3), for any objective j, the individual after the crossing is $U_{j,i,G} = \begin{cases} u_{j,i,G} & if\ rand_j[0,1) < CR \vee j = j_{rand} \\ x_{j,i,G} & otherwise \end{cases}$. To ensure that at least one gene in $u_{j,i,G}$ is passed on to the offspring, a gene position is selected for crossing randomly, that is, when $j = j_{rand}$, the rest of the loci are determined by $CR$ to be inherited from $u_{j,i,G}$ or $x_{j,i,G}$. $CR$ is the crossover probability, and its value can be set according to the specific problem. In this article, the value of $CR$ is 0.5.

Step (5). Differential selection: Select the individuals to join the offspring population S. By comparing the parent individual $x_{i,G}$ and intermediate individual $U_{i,G}$, select the individuals of the offspring population S, and the selection strategy is as follows:

1. Both solutions are infeasible, and if the parent individual dominates the intermediate individual in the solution space of constraint violation, then the parent individual is selected to enter the next generation, otherwise, the intermediate individual is selected to join the next generation.

2. If one of the solutions is feasible and the other is infeasible, then the feasible solution is selected to join the next generation.

3. Both solutions are feasible. If the parent individual dominates the intermediate individual, the parent individual will be selected to join the next generation, and vice versa; if the two individuals do not dominate each other, both solutions will join the next generation.

Step (6). Environment selection: Through step (5), the number of individuals entering the population S is between N and 2*N. In order to control the number of individuals in the population to N, the individual environment selection is required to select excellent individuals. The process is as follows:

1. The individuals in the offspring population S are sorted by using the fast non-dominated sorting strategy, and the first, second, ..., nth non-dominated fronts are obtained respectively.

2. Clear the parent population P, and set $remain = N$, $front$ is the number of individuals at the front. Obtain the first non-dominated front in step 1, if $(remain > 0)\ \&\&(remain \geq front)$, then the crowding distance of all individuals will be calculated and join to population P, $remain = remain - front$; then if $remain > 0$, we will obtain the second non-dominated front, and so on.

3. If $remain > 0$ and $front > remain$. Then, remove the poor individuals after calculating the crowding distance of the non-dominated front, and add excellent individuals to the population P, until the number of individuals in this population P is N.

If the end condition is satisfied, that is, $G \geq G_{max}$, the output population P is the Pareto approximate solution set; otherwise, turn to Step (3), G=G+1.

Step (7). The best compromise solution processing mechanism: The Euclidean distance between the individual in the optimal approximate solution set and the ideal point is calculated respectively, and the individual with the smallest distance is selected as the best compromise solution.

The calculation flow chart of BCS-GDE is shown in Fig.11:

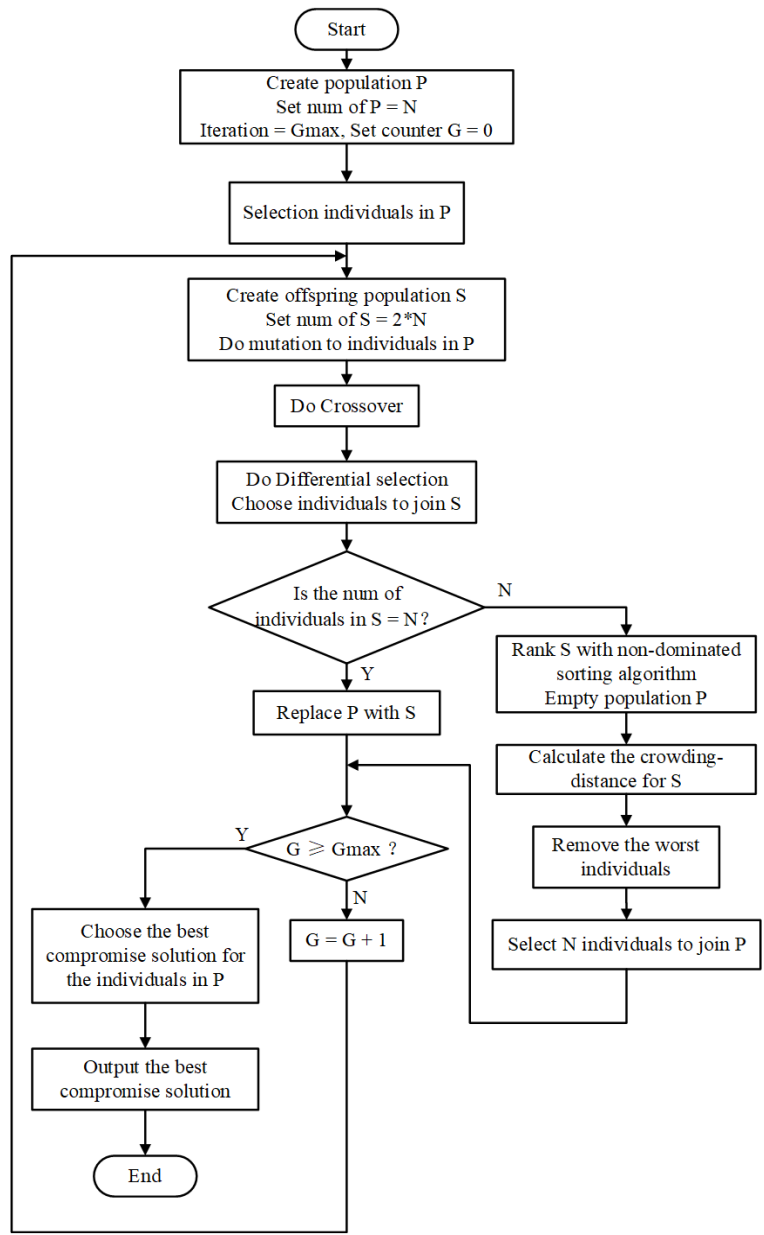

Fig 11 Flow chart of BCS-GDE

### 3.2 The choice strategy of the best compromise solution

Under the premise of minimizing the objective functions, the solution to the multi-objective CCHP model is a set of optimal solutions. These solutions dominate each other and cannot determine which is the optimal solution. The author adopts the best compromise solution selection strategy. By calculating the distance from each point to the ideal point, the point with the smallest distance is selected as the approximate optimal solution. It not only preserves various results for multi-objective problems but also selects a more suitable solution. In this paper, it is the optimal resource scheduling scheme. A schematic diagram of the distance calculation between the non-dominated solutions and the ideal point is shown in Fig.12.

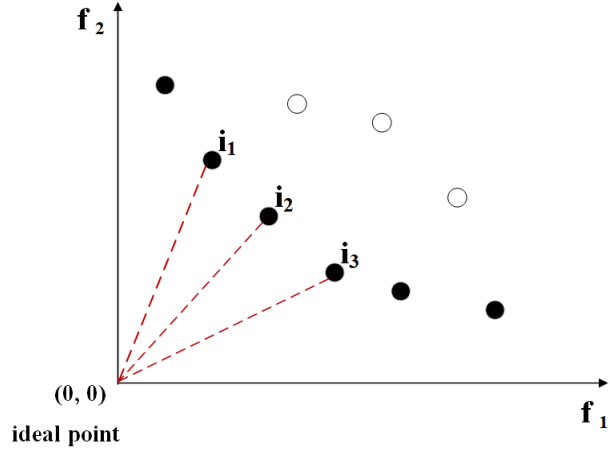

Fig 12 Schematic diagram of distance calculation

In Fig.12, $i_1$, $i_2$, and $i_3$ represent three non-dominated solutions, respectively, $f_1$ and $f_2$ represent two objective functions.

Generally speaking, the most ideal situation is that the three objective functions in the model reach the minimum values simultaneously, and the ideal point in the solution space is (0, 0, 0), but it is not likely to occur in practical problems. Therefore, in the process of choosing the best compromise solution, the Euclidean distances between the non-dominated solutions and the ideal point are calculated, the solution with minimum distance is chosen as the approximate optimal solution. The calculation method is as follows (41):

$$Dist(X,Y) = \sqrt{\sum_{i=1}^{n}(x_i - y_i)^2} \qquad (41)$$

where $X$ and $Y$ represent two n-dimensional vectors respectively, X $(x_1, x_2, ..., x_n)$, $Y(y_1, y_2, ..., y_n)$. In this paper, $Y(y_1, y_2, ..., y_n) = (0, 0, ..., 0)$.

## 4 Case study

In the simulation, the energy supply is provided for three types of buildings including hotels, offices, and residential buildings. The input data and parameter information are summarized in Section 4.1, and Section 4.2 analyzes different application scenarios in detail and analyzes the 24-hour energy demand of each building under different scenarios respectively.

### 4.1 Simulation settings

It is assumed that the virtual buildings used for simulation are all located in Xi'an, Shaanxi, China. There are four independent buildings in the hotel area with a total area of $202,768 m^2$, four independent buildings in the office area with a total area of $197,568 m^2$, and eight independent buildings in the residential area with a total area of $199,064 m^2$ [37]. In the CCHP system, the electricity price on the power grid and the price of natural gas are shown in Table 2. The electricity prices of every type of building are different in each power consumption period, as well as the price of natural gas, all the data are collected locally. In the case calculation, natural gas is used for PGU and boiler, the power grid, and PGU supply power.

Table 3 is the factor of converting primary energy to available energy, and every type of energy has different conversion factors. The CCHP system produces emissions when it works. Table 4 shows the carbon dioxide conversion factors of electric energy and natural gas. Table 5 shows the constraints of conversion and efficiency of CCHP in the process of energy output and transmission.

Table 2 Electricity and natural gas prices (Yuan/kWh).

|  | Electricity price | | Natural Gas price |
|---|---|---|---|
|  | Commercial buildings | Residential buildings |  |
| Average time | 0.87 | 0.5 | 0.22 |
| Peak load time | 1.305 | 0.65 | 0.22 |
| Low load time | 0.435 | 0.45 | 0.22 |

Table 3 Site-to-primary energy conversion factor [14].

| Fuel type | Conversion factor |
|---|---|
| Electricity | 3.336 |
| Natural Gas | 1.047 |

Table 4 Emission conversion factors for electricity and natural gas.

| Emission | Electricity(g/kWh) | Natural Gas(g/kWh) |
|---|---|---|
| $CO_2$ | 203.74 | 200 |

Table 5 Conversion and efficiency constraints [14].

|  | Symbol | Value |
|---|---|---|
| Fuel-to-electric-energy conversion parameter | $a$ | 2.67 |
| Fuel-to-electric-energy conversion parameter | $b$ | 11.43 |
| Fuel-to-thermal-energy conversion efficiency of PGU | $\eta_{pgu\_th}$ | 0.51 |
| Boiler efficiency | $\eta_{boiler}$ | 0.9 |
| Total efficiency of the cooling components | $\eta_{c\_comp}$ | 0.7 |
| Total efficiency of the heating components | $\eta_{h\_comp}$ | 0.85 |

The optimization models of the CCHP system have been established in Section 2, the process of the simulation will be addressed. To unify the simulation environment, all parameter settings are the same during the simulation, the parameters of BCS-GDE are listed in Table 6. The object of the BCS-GDE algorithm is population. Population is a group composed of several individuals, which represents a subset of the whole search space. Each individual in the population is a solution to the optimization problem, and the size of the population is the number of individuals.

Table 6 Parameters of BCS-GDE.

| Parameter | Value |
| --- | --- |
| **Population size** | 100 |
| **Maximum iterations** | 250 |
| **CR** | 0.5 |
| **F** | 0.5 |

## 4.2 Scenario analysis

In the simulation, we collected energy consumption in three real scenarios: hotels, offices, and residential buildings. The data are collected in summer, winter, and transition seasons respectively, and the average energy consumption in different seasons is shown in Fig.13. To observe the change in energy demand throughout the day, the energy demand is listed hourly.

In winter, less cooling is needed, and the demand for heating is the least in summer. While the demand has different rules according to buildings, the demand of the hotel is relatively large and stable, offices have a larger heating and cooling demand during normal working hours, residential buildings produce a certain demand according to residents' living habits. In transition seasons, there are both heating and cooling demands.

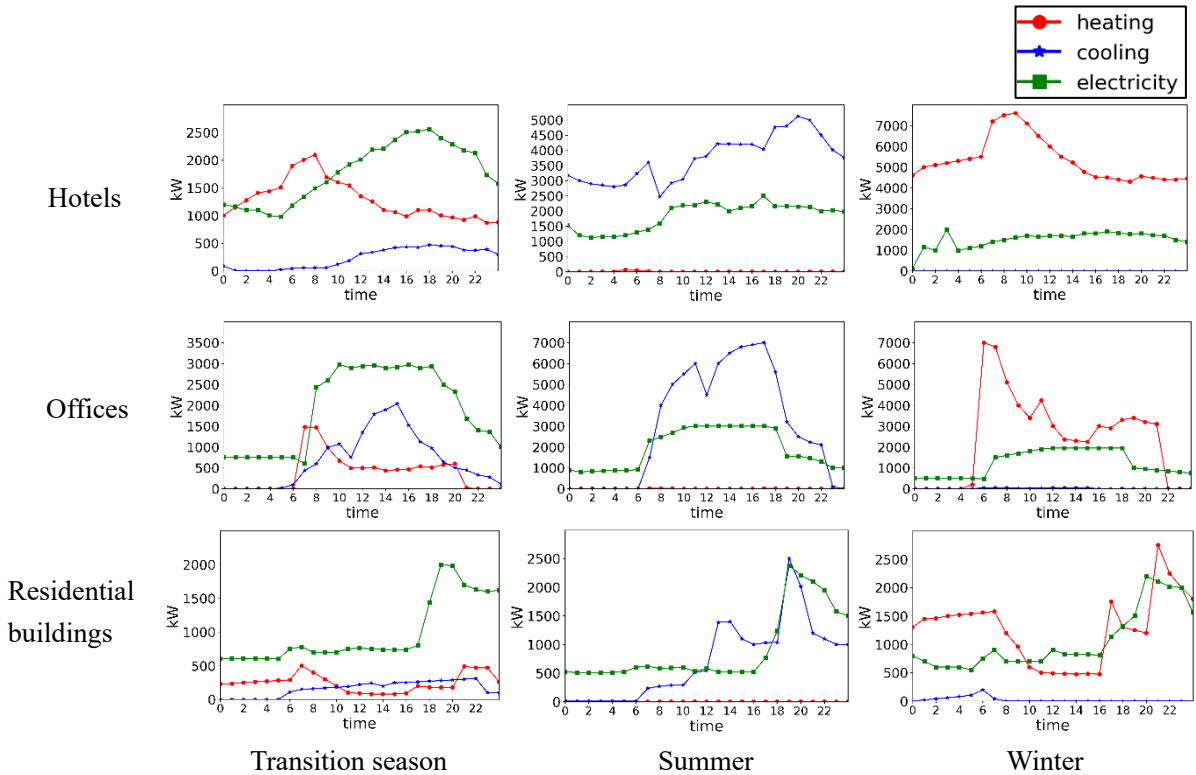

Fig 13 Daily energy load of buildings in the CCHP system

Table 7 shows the peak energy demand of various types of buildings.

Table 7 The electricity, cooling, and heating load of the buildings (kW).

|  | Hotel | Office | Residential buildings |
|---|---|---|---|
| **Electricity load** | 3070 | 3198 | 4166 |
| **Cooling load** | 5400 | 7056 | 6145 |
| **Heating load** | 7657 | 7050 | 7080 |

# 5 Results and discussion

The CCHP system models and BCS-GDE algorithm proposed in this paper are all implemented in jMetal 4.5 with JDK 1.8, and the computer environment is 2.8 GHz Intel Core i7, 8 GB RAM. The following two contents will be verified in this section: 1. The CCHP model proposed in this paper can effectively reduce economic costs, energy consumption, and pollutant emissions; 2. The BCS-GDE algorithm is more suitable for the optimization problem of energy scheduling than the classical algorithms, and the effect is remarkable. For verification content 1, simulation experiments are carried out in three scenarios to verify the advantages of the CCHP model proposed in this paper. For verification content 2, various algorithm comparison experiments are also carried out in jMetal. The CCHP model is optimized under the same environment and conditions, and the optimization results are compared.

## 5.1 Scenario 1- Transition season

The scenario in this section is in the transitional season, referring to spring and autumn. From Fig.13, it can be seen that in these seasons, the demand for power of all kinds of buildings is more than that of other seasons, and there is a demand for cooling and heating in this season; the demand for energy in offices fluctuates obviously in a day, while that in hotels and residential buildings is relatively flat.

### 5.1.1 Validation of the CCHP model

To verify the validity of the CCHP model, we first define the reference system. In the simulation, the reference system means the energy supply without the CCHP system. This means that the power of the reference system is purchased from the power grid, while the heating and cooling demand are supplied by using natural gas as the fuel through heating and cooling components. Thus, the economic cost, energy consumption, and pollutant emission of the reference system can be calculated directly. Three objective values of the CCHP system are optimized by the BCS-GDE algorithm. The three objective values of the reference system are shown in Table 8.

Table 8 Objective values for the reference system in the transition season

| Objective | Hotel | Office | Residential buildings |
|---|---|---|---|
| **Cost (Yuan)** | 60112.091 | 57047.784 | 21912.8702 |
| **PEC (kwh)** | 254635.1317 | 233241.9159 | 110227.8483 |
| **CDE (g)** | 29339770.02 | 23938795.7 | 10436697.04 |

Fig.14 states the change rate of economic cost, PEC, and CDE between the reference system and the CCHP model strategies, respectively. Case 1 represents the CCHP system with PGU and boiler, case 2 represents the model when PGU is shut down, case 3 represents the model when the boiler is shut down.

The vertical axis in the graph is the percentage reduction. This represents the reduction of the CCHP system relative to the reference system for each objective.

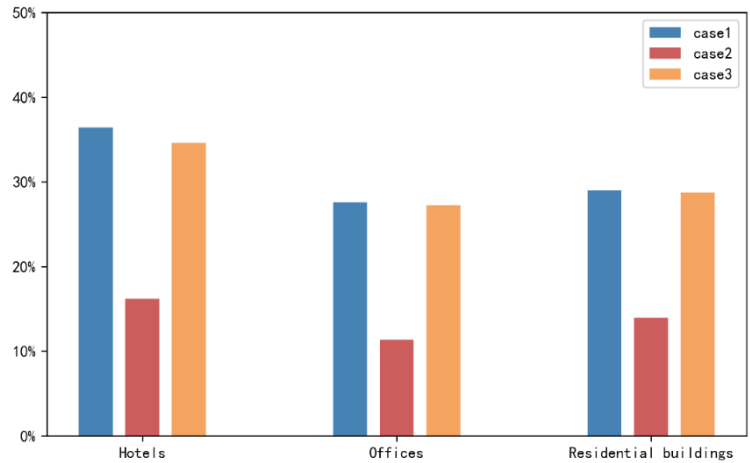

Fig 14 (a) Change rate of economic cost in scenario 1

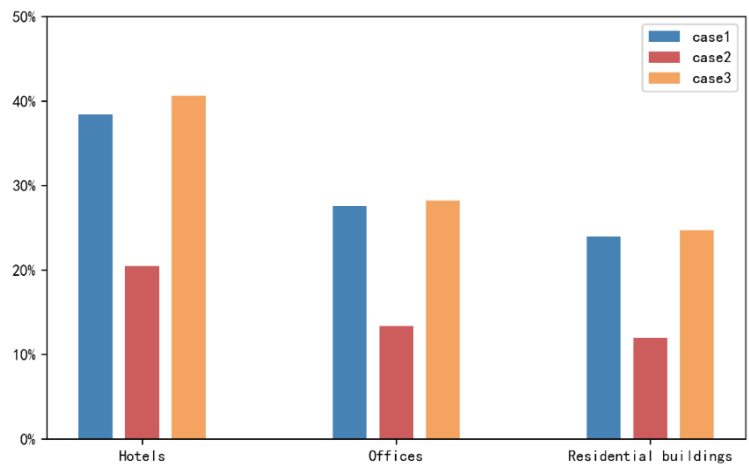

Fig 14 (b) Change rate of PEC in scenario 1

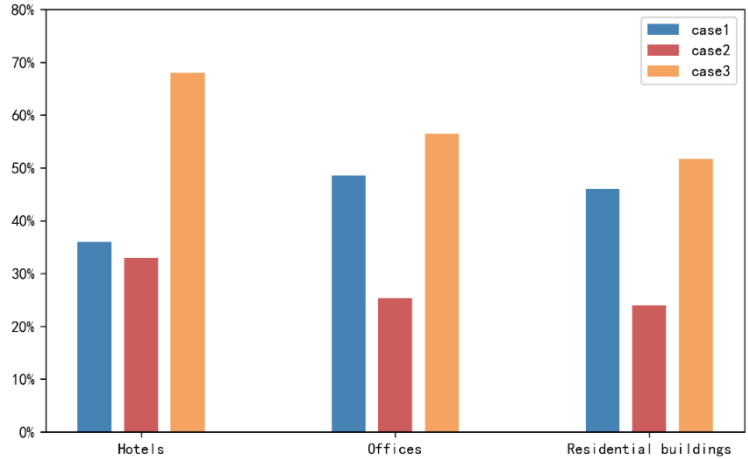

Fig 14 (c) Change rate of CDE in scenario 1

The authors verified the change in the objective value of three different types of buildings after using the CCHP system, respectively. Fig.14(a) shows the rate of decline in economic cost, hotels reduced costs by 36% under case 1 compared to the reference system, offices have the same decline in case1 and case3, both are 27.5%, residential buildings have a higher cost reduction rate under case 3 than the other cases at 29%. For the objective of PEC, case 3 has the most obvious reduction rate, saving more than 25% of primary energy consumption for the system. For CDE, case 3 has a change rate of 52% to 69% compared with the reference system, which can reduce more than half of the pollutant emissions of the system.

To sum up, the CCHP system proposed in this paper can effectively reduce cost, energy consumption, and pollutant emissions. Both in the objective of PEC and CDE, case 3 performance a better improvement than case 1, in the objective of economic cost, both of them have a similar result. In this scenario, case 3 will be accepted as the most suitable scheme to supply energy.

### 5.1.2 Verification of BSC-GDE algorithm

The BCS-GDE algorithm is used to solve the multi-objective optimization model and obtain the dispatching scheme of energy supply, the model is aiming at the minimum economic cost, PEC, and CDE. BCS-GDE algorithm obtains 100 non-dominated solutions according to the different energy demands of each day. In the simulation, the best compromise solution will be selected by using the rules in Section 3.2.

To verify the effectiveness and superiority of the proposed algorithm BCS-GDE in solving the CCHPEED, the simulation takes the peak energy demand of residential buildings in Table 7 as the rated energy demand. The Pareto approximate optimal solution calculated by the algorithm is shown in Fig. 15. The three axes of Fig.15 are three objective functions, respectively.

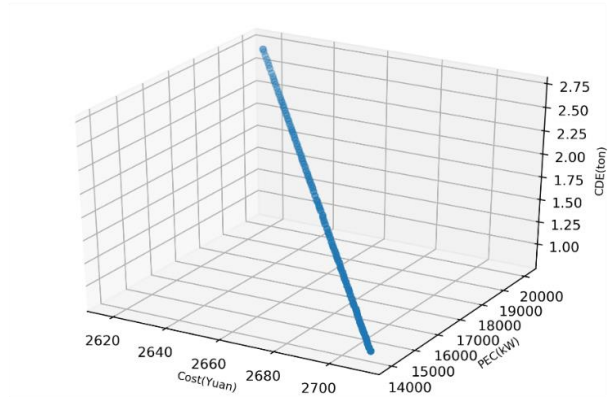

Fig 15 (a) Pareto approximation front of BCS-GDE

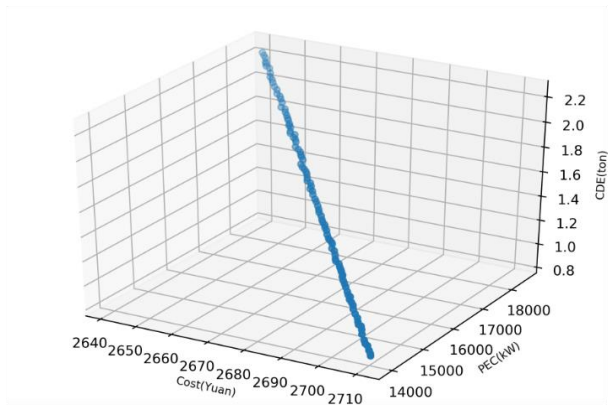

Fig 15 (b) Pareto approximation front of OMOPSO

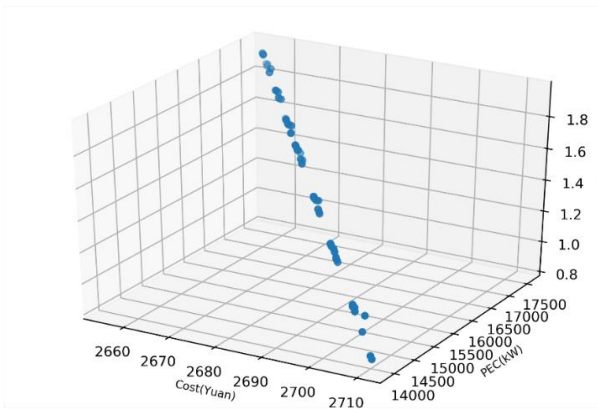

Fig 15 (c) Pareto approximation front of NSGA-II

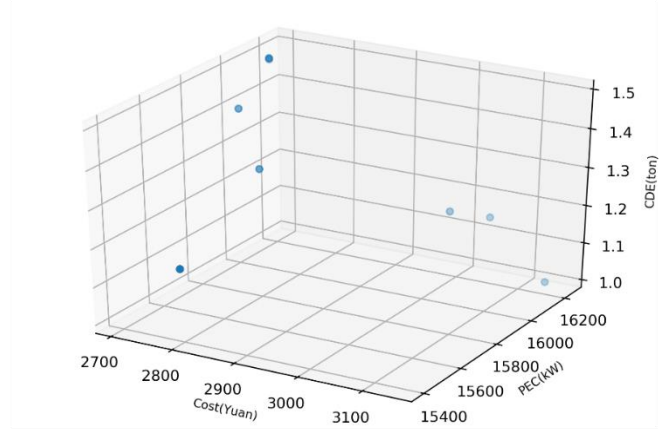

Fig 15 (d) Pareto approximation front of SPEA2

It can be seen from Fig.15 that in the same experimental environment, BCS-GDE can find the non-dominated solution with wide and evenly distributed, and the non-dominated solution found by OMOPSO has certain universality but poor convergence compared with BCS-GDE. The non-dominated solution obtained by NSGA-II has uneven distribution and few solutions, while SPEA2 can only find very few non-dominated solutions with poor convergence.

To compare algorithm performance objectively, Hypervolume (HV) and spread ($\triangle$) indicators are selected to evaluate the comprehensive performance and diversity of BCS-GDE and other algorithms in the simulation. To ensure the objective evaluation, each algorithm runs 20 times to obtain 20 indicator values respectively, and the maximum, minimum, and average values of each indicator value are shown in Table 9.

Table 9 Quality evaluation of OMOPSO, NSGA-II, SPEA2, and BCS-GDE

| Indicator | OMOPSO | | | NSGA-II | | | SPEA2 | | | BCS-GDE | | |
|---|---|---|---|---|---|---|---|---|---|---|---|---|
| | max | min | ave | max | min | ave | max | min | ave | max | min | ave |
| HV | 0.33 | 0.32 | 0.32 | 0.32 | 0.30 | 0.31 | 0.33 | 0.27 | 0.31 | 0.33 | 0.33 | **0.33** |
| $\triangle$ | 0.27 | 0.15 | 0.20 | 1.24 | 0.96 | 1.13 | 1.32 | 0.95 | 1.15 | 0.18 | 0.13 | **0.15** |

The larger the HV indicator value is, the better the comprehensive performance of the algorithm is, and the smaller the spread indicator value is, the more extensive and uniform the solution can be distributed on the Pareto front. According to Table 9, the HV value of BCS-GDE is 0.33, which is larger than that of the other algorithms. The spread indicator value of BCS-GDE is 0.15, which is the smallest compared with other algorithms. It shows that BCS-GDE not only has a more extensive and uniform distribution of non-dominated solutions, but also has better comprehensive performance compared with OMOPSO, NSGA-II, and SPEA2.

To verify the statistical results of the algorithms, the Wilcoxon signed-rank test is used for the performance difference of pair-wise comparison algorithms. The following hypothesis is proposed:

$H_0$: BCS-GDE has a significant improvement over the algorithm

Table 10 shows the statistical results, the p-value is considered to reject $H_0$ or not. In Table

10, all the p-values are less than the significance level α, and the hypothesis is accepted, $H_0$: BCS-GDE has a significant improvement over OMOPSO, NSGA-II, and SPEA2.

Table 10 Wilcoxon signed-rank test results with significance level α = 0.001.

| Methods | p-value (HV) | p-value (Spread) |
|---|---|---|
| **BCS-GDE vs OMOPSO** | 0.001 | 0.0001 |
| **BCS-GDE vs NSGA-II** | 0.0001 | 0.0001 |
| **BCS-GDE vs SPEA2** | 0.0001 | 0.0001 |

Table 11 lists the scheduling results of the algorithms to optimize the CCHP system. Among them, BCS-GDE has the smallest cost than the other three algorithms. At the same time, it has considerable PEC and CDE values.

Table 11 Energy dispatch results of residential buildings in transition seasons

| Methods | X1 | X2 | X3 | Cost (Yuan) | PEC (kwh) | CDE (g) |
|---|---|---|---|---|---|---|
| **OMOPSO** | 3392 | 2077 | 3.5 | 2663 | 17141 | 1803716 |
| **NSGA-II** | 3397 | 2063 | 0 | 2662 | 17113 | 1796219 |
| **SPEA2** | 3684 | 1350 | 68 | 2707 | 16149 | 1487583 |
| **BCS-GDE** | **3402** | **2051** | **0** | **2662** | **17095** | **1790792** |

## 5.2 Scenario 2- Summer

In summer, the demand for cooling and power supply of various buildings is significantly increased, while the demand for heating is almost zero. The energy demand of every building presents different characteristics, among which the energy demand of hotels is relatively stable in a day, and there are some fluctuations in residential buildings and offices.

### 5.2.1 Validation of the CCHP model

In the simulation experiments in this section, according to the energy demand in summer, the values of the reference system on the three objectives have been calculated, as shown in Table 12.

Table 12 Objective values for the reference system in the summer

| Objective | Hotel | Office | Residential buildings |
|---|---|---|---|
| **Cost (Yuan)** | 94920.4302 | 86371.611 | 25502.0162 |
| **PEC (kwh)** | 414696.7443 | 372217.4999 | 126940.6674 |
| **CDE (g)** | 59255686.56 | 50174058.5 | 14287279.72 |

In the face of the substantial growth of energy demand, the effectiveness of the CCHP models proposed in this paper is verified in the simulation. In Fig.16, the economic cost, PEC, and CDE change rate between the reference system and the strategies are shown respectively.

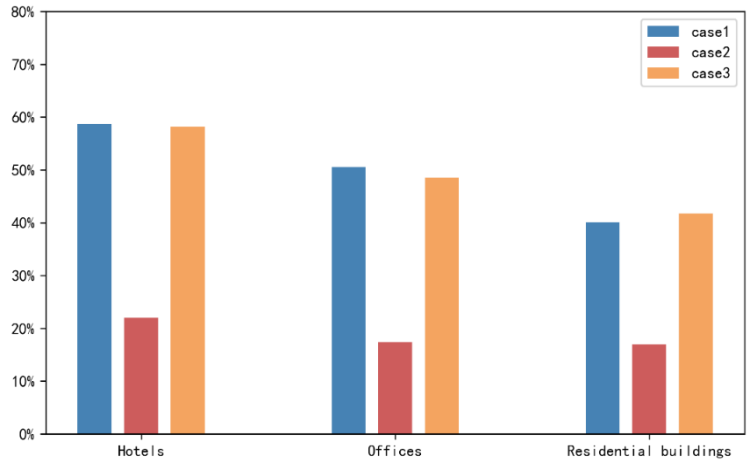

Fig 16 (a) Change rate of economic cost in scenario 2

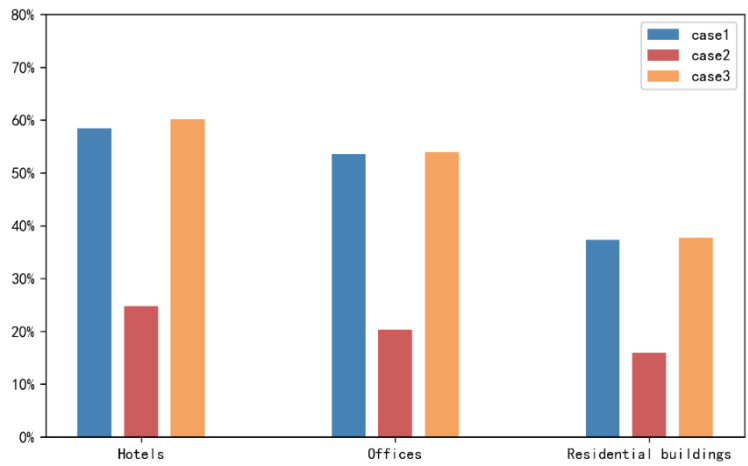

Fig 16 (b) Change rate of PEC in scenario 2

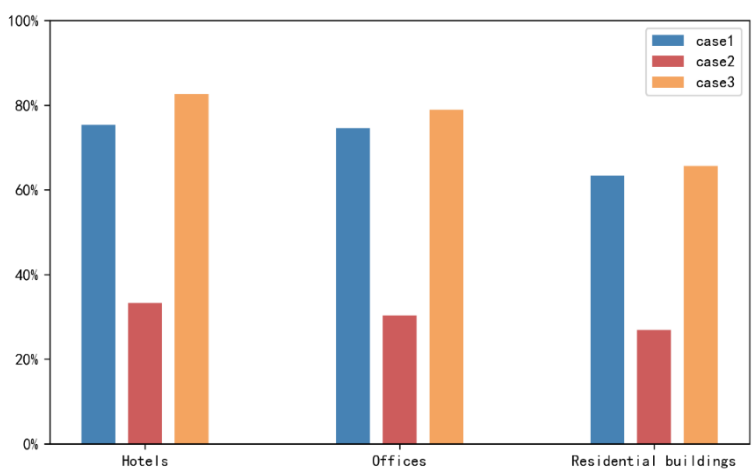

Fig 16 (c) Change rate of CDE in scenario 2

As can be seen from Fig.16, compared with the reference system, the CCHP models can significantly reduce the economic cost, PEC, and CDE, especially in the hotel scenario. For the objective of economic cost, case 1 has a maximum change rate of 59% in the hotel. For PEC, case 1 and case 3 reduce more than 50% of the primary energy consumption in hotels and offices and 38% in residential buildings. For the objective of CDE, the change rate of case 1 and case 3 is above 64%, and the maximum change rate of case 3 is 83% in the hotel scenario. Compared with the reference system, it can effectively and significantly reduce pollutant emissions.

In this scenario, case 2 with PGU off has the smallest improvement than the other scheme, for PGU provides both power and heating, as seen in Fig.8. For the simulation, the performance of PGU is better than that of the boiler. Case 3 can be chosen for energy supply for hotels, and both case 1 and case 3 can be used for energy supply for offices and residential buildings.

### 5.2.2 Verification of BSC-GDE algorithm

The peak energy demand of hotels in Table 7 is used to verify the performance of BCS-GDE. Fig.17 shows the Pareto approximation fronts obtained by BCS-GDE and three classical algorithms to optimize the CCHP model.

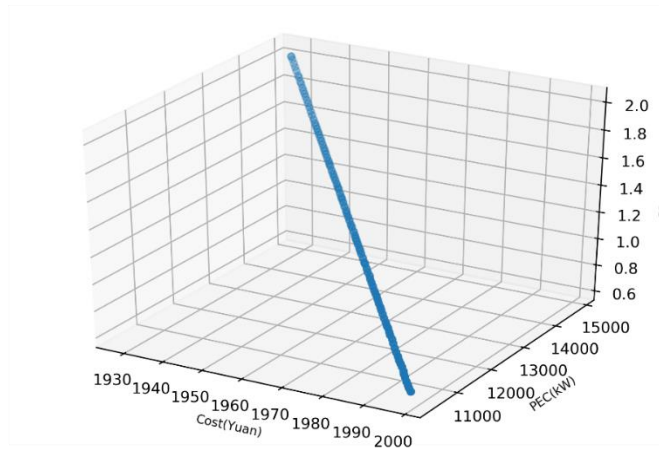

Fig 17 (a) Pareto approximation front of BCS-GDE

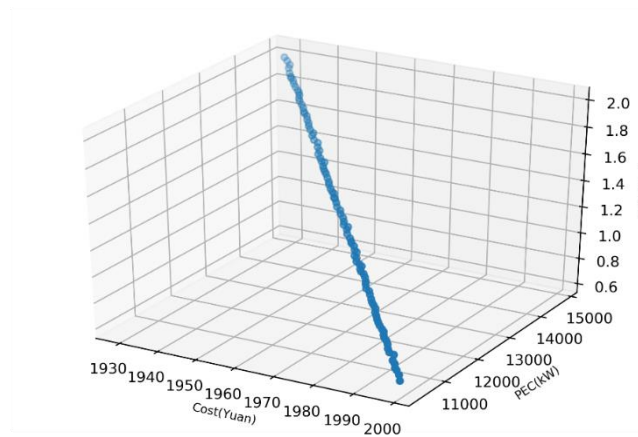

Fig 17 (b) Pareto approximation front of OMOPSO

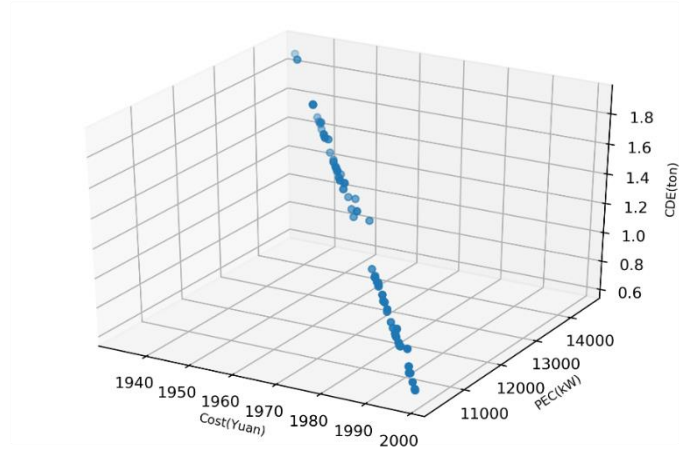

Fig 17 (c) Pareto approximation front of NSGA-II

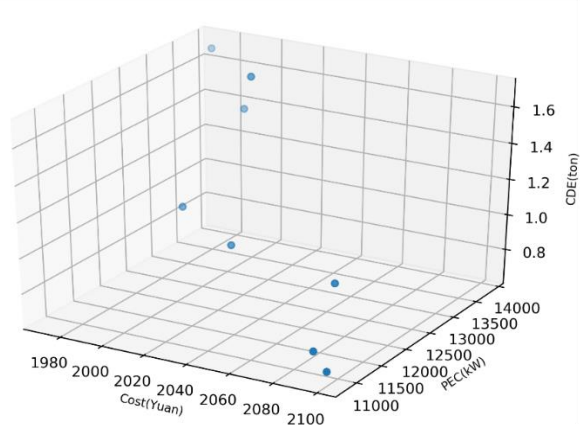

Fig 17 (d) Pareto approximation front of SPEA2

It can be seen from Fig.17 that the non-dominated solution obtained by BCS-GDE has better distribution and convergence than other algorithms, the solution obtained by NSGA-II is not evenly distributed, while the number of feasible solutions of SPEA2 in the same experimental environment is very small, and the quality of the solution is poor.

To show the performance of the algorithm more intuitively, HV and spread indicators are used to evaluate the algorithms, as shown in Table 13.

Table 13 Quality evaluation of OMOPSO, NSGA-II, SPEA2, and BCS-GDE.

| Indicator | OMOPSO | | | NSGA-II | | | SPEA2 | | | BCS-GDE | | |
| --- | --- | --- | --- | --- | --- | --- | --- | --- | --- | --- | --- | --- |
| | max | min | ave | max | min | ave | max | min | ave | max | min | ave |
| HV | 0.33 | 0.31 | 0.32 | 0.31 | 0.28 | 0.30 | 0.49 | 0.17 | 0.29 | 0.33 | 0.33 | **0.33** |
| Δ | 0.25 | 0.17 | 0.20 | 1.22 | 0.94 | 1.09 | 1.28 | 0.71 | 1.00 | 0.16 | 0.12 | **0.15** |

It can be seen from Table 13 that BCS-GDE has a maximum HV value of 0.33 and the minimum spread value of 0.15 compared with OMOPSO, NSGA-II, and SPEA2. BCS-GDE algorithm has a better performance than other algorithms in different energy demands.

The results of the Wilcoxon test are as Table14, p-values are less than the significance level α, the hypothesis can be accepted as BCS-GDE has a significant improvement over OMOPSO, NSGA-II, and SPEA2.

Table 14 Wilcoxon signed-rank test results with significance level α = 0.5.

| Methods | p-value (HV) | p-value (Spread) |
|---|---|---|
| **BCS-GDE vs OMOPSO** | 0.001 | 0.0001 |
| **BCS-GDE vs NAGA-II** | 0.0001 | 0.0001 |
| **BCS-GDE vs SPEA2** | 0.1 | 0.0001 |

The scheduling results of the algorithms to optimize the CCHP system are listed in Table 15. From Table 15, the cost, PEC, and CDE of BCS-GDE are smallest than the results of other algorithms. The algorithm proposed in this paper optimizes the CCHP model to obtain the optimal scheduling scheme, which has a significant impact on the three objectives.

Table 15 Energy dispatch results of hotels in the summer

| Methods | X1 | X2 | X3 | Cost (Yuan) | PEC (kwh) | CDE (g) |
|---|---|---|---|---|---|---|
| **OMOPSO** | 2144 | 2483 | 1 | 1940 | 14109 | 1765565 |
| **NSGA-II** | 2495 | 1548 | 0 | 1962 | 12665 | 1337617 |
| **SPEA2** | 2440 | 1823 | 61 | 2001 | 13315 | 1485335 |
| **BCS-GDE** | **2859** | **575** | **0** | **1961** | **12662** | **1337692** |

### 5.3 Scenario 3- Winter

The scenario in this section is in winter. As shown in Fig.13, the demand for heat supply is more than that for cold supply. Besides, offices have obvious fluctuations in heating and power demand, and the energy demand of the hotel is relatively stable.

#### 5.3.1 Validation of the CCHP model

In this section, the three objective values of each building group under the reference system in winter are calculated according to the energy demand, as shown in Table 16.

Table 16 Objective values for the reference system in the winter

| Objective | Hotel | Office | Residential buildings |
|---|---|---|---|
| **Cost (Yuan)** | 111155.6886 | 62121.0366 | 36157.3978 |
| **PEC (kwh)** | 498738.1571 | 270807.5519 | 178463.034 |
| **CDE (g)** | 78961324.8 | 38347759.8 | 22677360.98 |

Fig.18 shows the change rate of economic cost, PEC, and CDE between the reference system and the CCHP models.

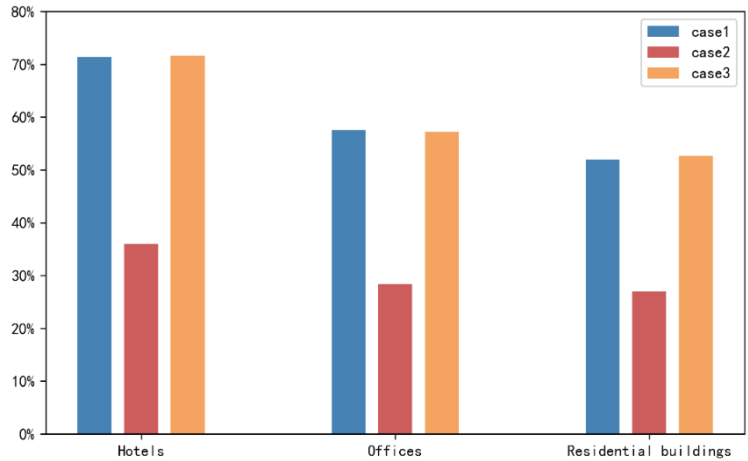

Fig 18 (a) Change rate of economic cost in scenario 3

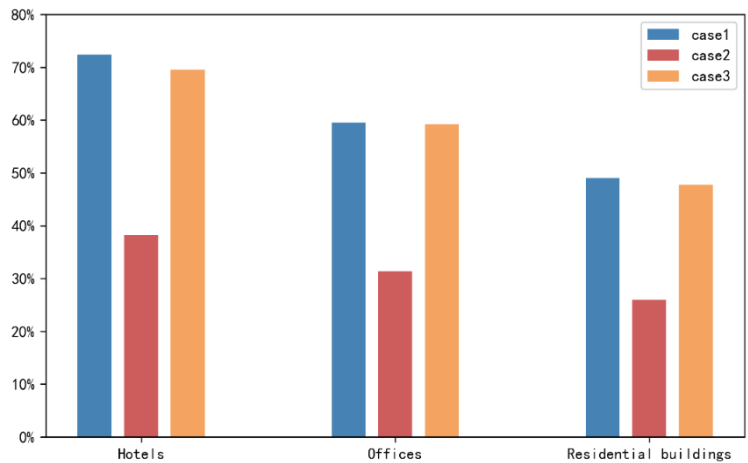

Fig 18 (b) Change rate of PEC in scenario 3

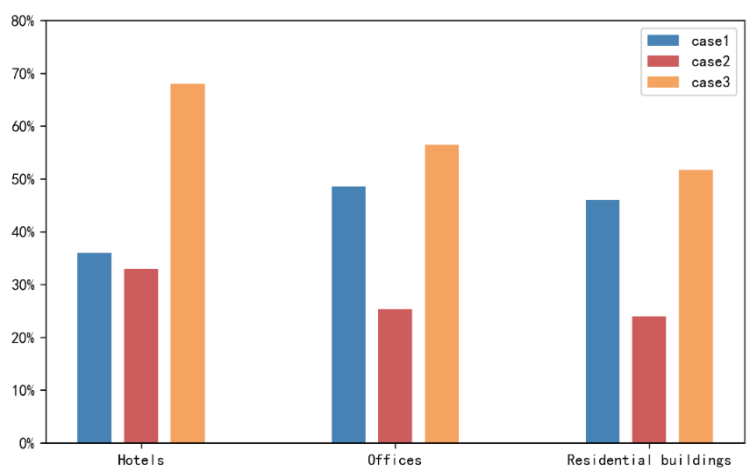

Fig 18 (c) Change rate of CDE in scenario 3

There are obvious differences between the three strategies relative to the reference system. The change rate of case 1 and case 3 on the objectives is above 48%, the maximum change rate of case 1 and case 3 in the economic cost objective is 71%, and the maximum change rate of case 3 is 72%. In the PEC objective, the maximum change rate of case 1 is 73%, and case 3 is 70%. The maximum change rate of case 1 and case 3 is 86% and 88% respectively in the CDE objective. The change rate of case 2 in the three objectives is more than 26%, and the maximum change rate is 46%.

The simulation results address that both case1 and case3 performance significantly reduce in the objective of economic cost, PEC, and CDE. Considering synthetically multiple factors, case3 can be used to supply energy for offices, both case1 and case3 can be used for hotels and residential buildings in winter.

### 5.3.2 Verification of BSC-GDE algorithm

To verify the universality of BCS-GDE in the CCHP system, the peak load of the office in Table 7 is used as the rated demand in the experiment, and BCS-GDE is used for calculation and compared with other algorithms. The simulation results are shown in Fig.19.

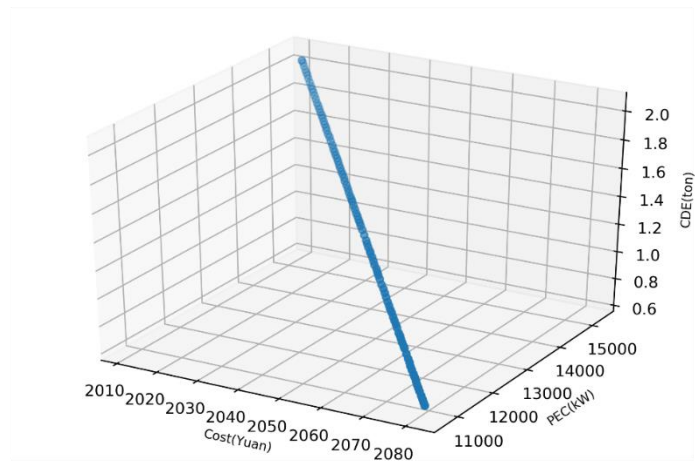

Fig 19 (a) Pareto approximation front of BCS-GDE

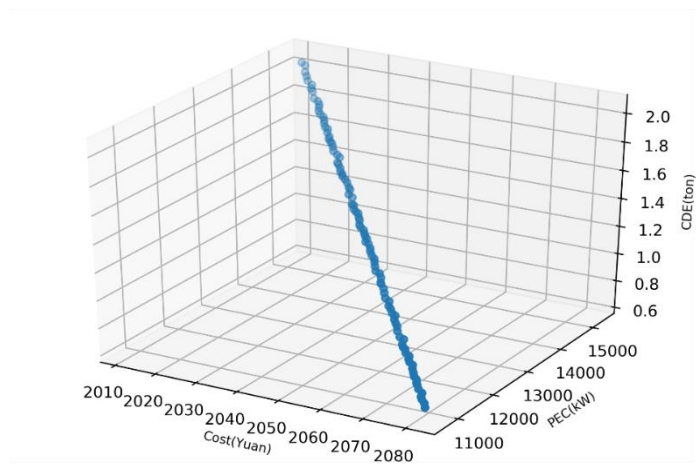

Fig 19 (b) Pareto approximation front of OMOPSO

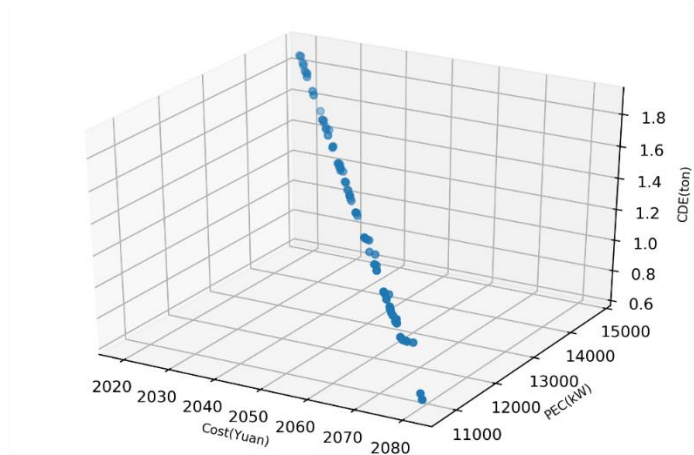

Fig 19 (c) Pareto approximation front of NSGA-II

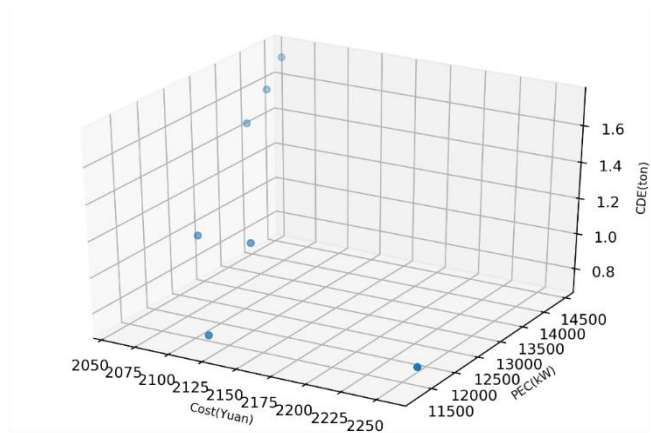

Fig 19 (d) Pareto approximation front of SPEA2

It can be seen from Fig.19 that the non-dominated solution obtained by BCS-GDE is widely distributed and uniform, and can also converge well. The non-dominated solution obtained by OMOPSO has a certain distribution, but its convergence is poor. The solution obtained by NSGA-II is not evenly distributed and has poor convergence. SPEA2 finds few non-dominated solutions to this problem, and the quality of the solutions is very poor.

Table 17 shows the evaluation results of the algorithms using the HV indicator and spread indicator.

Table 17 Quality evaluation of OMOPSO, NSGA-II, SPEA2, and BCS-GDE.

| Indicator | OMOPSO | | | NSGA-II | | | SPEA2 | | | BCS-GDE | | |
|---|---|---|---|---|---|---|---|---|---|---|---|---|
| | max | min | ave | max | min | ave | max | min | ave | max | min | ave |
| **HV** | 0.33 | 0.21 | 0.32 | 0.31 | 0.28 | 0.30 | 0.32 | 0.28 | 0.30 | 0.33 | 0.33 | **0.33** |
| Δ | 0.23 | 0.16 | 0.20 | 1.22 | 0.98 | 1.10 | 1.34 | 1.07 | 1.20 | 0.17 | 0.12 | **0.15** |

As shown in Table 17, the HV indicator value of BCS-GDE is 0.33, and the spread indicator value is 0.15. Compared with OMOPSO, NSGA-II, and SPEA2, BCS-GDE has the maximum HV value and the minimum spread value. It shows that BCS-GDE not only has a wide and uniform solution distribution but also has a high-quality comprehensive performance. Therefore, BCS-GDE is more suitable for solving CCHP than other algorithms, and it has obvious advantages in the process of optimization.

In Table 18, the p-values of the Wilcoxon test are less than the significance level α, we can accept the hypothesis that BCS-GDE has a significant improvement over the other three algorithms. In the simulation, BCS-GDE addresses better comprehensive performance than other algorithms.

Table 18 Wilcoxon signed-rank test results with significance level α = 0.0001.

| Methods | p-value (HV) | p-value (Spread) |
|---|---|---|
| **BCS-GDE vs OMOPSO** | 0.0001 | 0.0001 |
| **BCS-GDE vs NAGA-II** | 0.0001 | 0.0001 |
| **BCS-GDE vs SPEA2** | 0.0001 | 0.0001 |

Table 19 lists the optimization results of each algorithm for energy dispatching of offices. It can be seen that compared with other algorithms, BCS-GDE exhibits good performance on all three objectives, reducing cost, energy consumption, and pollutant emissions.

Table 19 Energy dispatch results of residential offices in winter

| Methods | X1 | X2 | X3 | Cost (Yuan) | PEC (kwh) | CDE (g) |
|---|---|---|---|---|---|---|
| **OMOPSO** | 2493 | 1553 | 0 | 1962 | 12669 | 1339419 |
| **NSGA-II** | 2495 | 1549 | 0 | 1963 | 12665 | 1337617 |
| **SPEA2** | 2352 | 2067 | 51 | 1995 | 13692 | 1595820 |
| **BCS-GDE** | **2859** | **575** | **0** | **1961** | **12663** | **1337692** |

# 6  Conclusion

This paper proposes an energy-optimized CCHP model for three types of buildings: hotels, offices, and residential buildings. In order to adapt to different energy structures, this paper constructs another two model strategies: shutting down the PGU or shutting down the boiler. For optimal energy dispatch, the BCS-GDE algorithm was proposed for the first time, the following are the conclusions that can be drawn.

(1) The CCHP model established in this paper can greatly reduce the economic cost, primary energy consumption, and carbon dioxide emissions. In terms of cost, PEC, and CDE objectives, the model can save 88%, 73%, and 72% of the reference system at most.

(2) In the simulation, the performance of algorithms is compared and analyzed, BCS-GDE has a confidence level of more than 95%, which is significantly improved than other algorithms. Through objective evaluation, the comprehensive performance indicator (HV) value of BCS-GDE is 0.3, and the convergence indicator (Δ) value is 0.15. This shows that it has better-optimized performance compared to other algorithms.

(3) To verify the validity of the model and algorithm, this paper provides the 24-hour energy demand of three types of buildings in the transition season, summer and winter, respectively. In the verification of the algorithm, using the multi-objective evaluation indicators can compare the performance of the algorithm more intuitively.

In future work, we will give more consideration to the use of clean energy, such as solar, wind, natural gas, and fuel cells.

# 7 Acknowledgement

The work is supported by the National Natural Science Foundation of China (Grant No. 61876138), the Key R & D Project of Shaanxi Province (2020GY-010), the Industrial Research Project of Xi'an (2019218114GXRC017CG018-GXYD17.10), and the Special Fund for Key Discipline Construction of General Institutions of Higher Learning from Shaanxi Province.

**NOMENCLATURE**

**Acronyms**

| | |
|---|---|
| CHP | combined heat and power |
| CCHP | combined cooling heating and power |
| CCHPEED | combined cooling heating and power economic emission dispatch |
| LP | linear programming |
| MILP | mixed integer linear programming |
| MINLP | nonlinear mixed integer linear programming |
| PSO | particle swarm optimization |
| DMOPs | dynamic multi-objective optimization problems |
| GA | genetic algorithm |
| NSGA-II | non-dominated sorting genetic algorithm-II |
| BCS-GDE | generalized difference algorithm with the best compromise solution processing mechanism |
| DE | differential evolution |
| GDE3 | the third evolution step of generalized differential evolution |
| SPEA2 | strength Pareto evolutionary algorithm 2 |
| OMOPSO | optimised multi-objective particle swarm optimization |
| PEC | primary energy consumption |
| CDE | carbon dioxide emissions |
| PG | power grid |
| PGU | power generation units |

**Symbols**

| | |
|---|---|
| $E_{grid}$ | electricity from the grid |
| $F_{pgu}$ | PGU fuel energy consumption |
| $F_{boiler}$ | boiler fuel energy consumption |
| $E_{pgu}$ | electricity from PGU |
| $Q_{rcv}$ | recovered waste heat from the PGU |
| $Q_{boiler}$ | thermal energy produced by the boiler |
| $E_{facility}$ | electric energy provided to the buildings |
| $Q_{th\_c}$ | thermal energy produced by the cooling components |
| $Q_{th\_h}$ | thermal energy produced by the heating components |
| $Q_c$ | cooling energy for the buildings |
| $Q_h$ | heating energy for the buildings |
| $E_{excess}$ | excess electric energy produced by the PGU |
| $E_{loss\_c}$ | energy loss of the CCHP cooling components |
| $E_{loss\_h}$ | energy loss of the CCHP heating components |
| $E_{loss\_pgu}$ | energy loss of the PGU |
| $E_{loss\_boiler}$ | energy loss of the boiler |
| $E_{loss\_total}$ | total energy loss |
| $E_d$ | electric energy demand |
| $Q_{c\_d}$ | cooling energy demand |
| $Q_{h\_d}$ | heating energy demand |
| $f_{cost}$ | the total cost of the system |
| $f_{PEC}$ | the primary energy consumption |
| $f_{CDE}$ | the carbon dioxide emissions |
| $C_{el}$ | the cost of purchasing 1kwh electricity |
| $C_{f_{pgu}}$ | the fuel cost of generating 1kwh energy in PGU |
| $C_{f_{boiler}}$ | the fuel cost of generating 1kwh energy in the boiler |
| $\eta_{boiler}$ | the energy conversion efficiency of the boiler |
| $ECF_{PEC}$ | the conversion factor of primary energy for purchasing power |
| $FCF_{PEC\_pgu}$ | the primary energy conversion factor of fuel used in PGU |
| $FCF_{PEC\_boiler}$ | the primary energy conversion factor of fuel used in the boiler |
| $a, b$ | fuel electric energy conversion parameters |
| $ECF_{CDE}$ | the carbon dioxide emission conversion factor of power |
| $FCF_{CDE\_pgu}$ | the emission conversion factor of PGU fuel |
| $FCF_{CDE\_boiler}$ | the emission conversion factor of boiler fuel |
| $\eta_{pgu\_th}$ | the conversion efficiency of PGU |
| $\eta_{boiler}$ | the conversion efficiency of the boiler |
| $\eta_{cool\_comp}$ | the conversion efficiency of the cooling component |
| $\eta_{heat\_comp}$ | the conversion efficiency of the heating component |
| $x_j^L, x_j^U$ | the lower and upper bounds of the individual x on the jth objective |
| $F$ | mutation operator |
| $CR$ | crossover probability |
| $X, Y$ | n-dimensional vectors |